\def\ieeefinal{1}

\ifnum\ieeefinal=1 %
\documentclass[times, 10pt,twocolumn]{article}
\usepackage{times}
\usepackage{hyperref}
\else %
\documentclass[11pt,letterpaper]{article}
\usepackage{hyperref}
\usepackage{fullpage}
\fi

\usepackage{amsmath, amsthm, amssymb}
\usepackage[dvips]{graphics}
\usepackage{geometry}
\newcommand{\bra}[1]{\langle #1|}
\newcommand{\ket}[1]{|#1\rangle}

\newcommand{\acc}{\mathsf{ac}}
\newcommand{\rej}{\mathsf{re}}

\newtheorem{thm}{Theorem}[section]

\newtheorem{lem}[thm]{Lemma}
\theoremstyle{remark}

\theoremstyle{definition}

\ifnum\ieeefinal=1 
\newenvironment{Spaceonleft}
{\begin{list}{}{\setlength{\leftmargin}{0pt}}
         \item[]}
  {\end{list}}
\newenvironment{myprot}{
\parindent 0pt
\rule{\columnwidth}{1pt}\begin{Spaceonleft}}
{\ignorespaces\end{Spaceonleft}\vspace{-12pt}\rule{\columnwidth}{1pt}
}
\else 
\newenvironment{Spaceonleft}
{\begin{list}{}{\setlength{\leftmargin}{18pt}}
         \item[]}
  {\end{list}}
\newenvironment{myprot}{
\parindent 0pt
\rule{\textwidth}{1pt}\begin{Spaceonleft}}
{\ignorespaces\end{Spaceonleft}\vspace{-12pt}\rule{\textwidth}{1pt}
}
\fi 

\usepackage{epsfig}

\begin{document}
\title{Secure Multiparty Quantum Computation with (Only) a Strict Honest
Majority}

\author{Michael Ben-Or\\ The Hebrew University\\ benor@cs.huji.ac.il \and
Claude Cr\'epeau \\  McGill University \\
crepeau@cs.mcgill.ca \and
Daniel Gottesman \\ Perimeter Institute for Theoretical Physics \\
dgottesman@perimeterinstitute.ca \and Avinatan Hassidim
\\ The Hebrew University
\\ avinatanh@gmail.com
\and Adam Smith\\ Weizmann Institute of
Science\\
adam.smith@weizmann.ac.il}


\maketitle

\begin{abstract}
Secret sharing and multiparty computation (also called ``secure
function evaluation'') are fundamental primitives in modern
cryptography, allowing a group of mutually distrustful players to
perform correct, distributed computations under the sole
assumption that some number of them will follow the protocol
honestly. This paper investigates how much trust is necessary --
that is, how many players must remain honest -- in order for
distributed quantum computations to be possible.

We present a verifiable quantum secret sharing (VQSS) protocol,
and a general secure multiparty quantum computation (MPQC)
protocol, which can tolerate any $\lfloor \frac{n-1}{2}\rfloor$
cheaters among $n$ players. Previous protocols for these tasks
tolerated $\lfloor \frac{n-1}{4}\rfloor$ and $\lfloor
\frac{n-1}{6}\rfloor$ cheaters, respectively. The threshold we
achieve is tight --- even in the classical case, ``fair''
multiparty computation is not possible if any set of $n/2$ players
can cheat.

Our protocols rely on approximate quantum error-correcting codes,
which can tolerate a larger fraction of errors than traditional,
exact codes. We introduce new families of authentication schemes
and approximate codes tailored to the needs of our protocols, as
well as new state purification techniques along the lines of those
used in fault-tolerant quantum circuits.
\end{abstract}

\if\ieeefinal=0
\newpage
\pagenumbering{arabic} %
\fi

\section{Introduction}\label{sec:intro}

Secure multiparty computation has been studied extensively in the
classical setting (see \cite{Gol97} for a survey) and was extended
to the quantum setting by \cite{CGS02}. A {\em secure quantum
multiparty protocol} (or {\em secure function evaluation}) allows
$n$ participants $P_1,\ldots,P_n$ to compute an $n$ input quantum
circuit where each player $P_i$ is responsible for providing one
of the input states. The output of the circuit is broken into $n$
components $H_1 \otimes \ldots \otimes H_n$ and $P_i$ receives the
output $H_i$. Note that the inputs are arbitrary (possibly
entangled) quantum states and each player simply has his input in
his possession --- he does not need to know its classical
description. Informally we wish to achieve the same functionality
as if each player were to hand his input to a trusted third party
who would evaluate the circuit and distribute the outputs.
Moreover we wish to do so even when up to $t$ players are faulty.

In the quantum setting it seemed at first that the best one could
hope for is to tolerate $t < n/4$ faulty players simply because
(exact) quantum error correcting codes (QECC) cannot recover from
more errors. Indeed the best previously known verifiable quantum
secret sharing protocol can tolerate $t<n/4$ faulty players, and
the best secure quantum multiparty protocol tolerates only $t<n/6$
faults \cite{CGS02}\footnote{The preliminary version of
\cite{CGS02} claims that the $n/4$ bound is tight for VQSS;
however, the bound holds only for errorless protocols. See
\cite{CGS05} for a discussion.}. However, approximate QECCs exist
\cite{CGS05} that can recover (with high probability) from the
corruption of $t < n/2$ shares, and their discovery paved the way
to this paper.

\smallskip
\noindent{\bf Main Result}
{\em Assuming pairwise quantum channels and a classical broadcast
channel between $n$ players, there exists a universally
composable, 
statistically secure multiparty quantum computation protocol, that
tolerates an adaptive adversary controlling up to $t < n/2$ faulty
players. The complexity of the protocol is polynomial in the
number of players and the size of the circuit.}

\smallskip

%
%
\noindent Note: Tolerating $t \ge n/2$ faulty players is not
possible in our model without computational assumptions. This
follows, for example, from the impossibility of unconditionally
secure quantum bit commitment for two players (see \cite{DKSW06}
for a recent discussion).


\medskip  In our setting, universally composable classical secure
multiparty computation is possible \cite{RB89,Can01,KLT06} and,
crucially, the proofs of composability hold even in the quantum UC
model \cite{BM04}.
One strategy we use extensively is to reduce the quantum
multiparty computation to a secure computation on classical keys.

\paragraph{Protocol Overview}
\newcommand{\secref}[1]{Section~\ref{sec:#1}}

Our protocol follows the basic ``share-and-compute" paradigm of
classical distributed protocols. Players use a quantum version of
{\em verifiable secret sharing} (called VQSS) to distribute an
encoding of their input. They then perform the circuit, gate by
gate, on these encoded values. The circuit's outputs are then sent
to appropriate players by opening the VQSS.

The structure of the verifiable sharing is similar to classical
protocols, with several important differences.  First, the
encoding used in the VQSS combines error-correction and
authentication in a novel way. Authenticating quantum data
requires encryption, and new tools are required to manipulate
encrypted data. In particular, we can often push the complexity of
the quantum computation into a classical computation on the
authentication (or encryption) keys -- this computation can be
performed with classical tools. One of our contributions is a
family of {\em self-dual} authentication codes, which make these
classical computations simpler to represent.

Second, the error-correcting code underlying the protocols is an
``approximate" code which tolerates any $t<n/2$ errors, similar to
that in~\cite{CGS05}. In order to perform a full (dense) set of
operations fault-tolerantly on these encodings, we develop a
procedure for purifying encodings of ``Toffoli states'', whose
creation is sufficient to perform a Toffoli gate.

Finally, because quantum data cannot be cloned, the dealer does
not use his actually data in the sharing protocol until after the
sharing is successfully completed -- he then inputs his data via
teleportation.

\paragraph{This Abstract}

The basic authentication code, and operations required on
authenticated data, are described in \secref{quanauth}.
Sections~\ref{sec:shareEpr}, \ref{sec:WQSS} and \ref{sec:VQSS}
build up the pieces necessary for VQSS. \secref{shareEpr} extends
the authentication scheme to {\em verified state authentication},
in which a dealer can prove to another player that a message is
correctly authenticated via a key shared among all players. The
next step is {\em weak} VQSS, which plays the role of a classical
commitment scheme: the dealer shares a state in such a way that he
cannot change it, but may refuse to open it at a later stage
(\secref{WQSS}). \secref{VQSS} explains the final layer needed for
VQSS. Finally, Sections~\ref{sec:MPQC} and \ref{sec:simul} give
the purification procedure for Toffoli states and a sketch of the
simulation argument showing that the whole protocol is secure.

\ifnum\ieeefinal=0
\paragraph{Simulation}

\nopagebreak We have taken extra care to guarantee that our
protocols will be universally composable. The protocols we present
are quite involved and might not have optimal complexity. They do
have one surprising proprety --- the simulation (section 8)
required for their correctness proof is simple and
straightforward.

\section{Preliminaries}\label{sec:prelim}
\fi

\paragraph{The Network Model}

We assume a \emph{synchronous} network (with rushing) in which
pairwise secure quantum channels exist between any two players,
and a classical broadcast channel connects all players.

\paragraph{The Adversary}

We assume that the adversary is computationally unbounded, and
that she fully coordinates the actions of all faulty players. She
may corrupt players adaptively during the course of the protocol.
There are two limitations to this: (1) At most $t < \frac{n}{2}$
players may become faulty during the course of the computation;
(2) the adversary has access only to the information of the
corrupted players she currently controls. We call the non-faulty
players {\em honest}.

All protocols we present have a success probability exponentially
close to one (also called an exponentially good probability) in
some security parameter\footnote{As proved in \cite{CGS02},
Verifiable Quantum Secret Sharing and Multiparty Quantum
Computation with more than $n/4$ faulty players must have some
probability of failure. Therefore, the Multiparty Computation
protocol we present will have an exponentially small probability
of failure in some security parameter. Let $S$ be the value of
this parameter, and $C$ the size of the quantum circuit we want to
securely evaluate. Our algorithms will be polynomial $n$, $S$ and
$C$. Setting a security parameter $s=S+n^2+\log C +O(1)$ in all
our subprotocols is sufficient to guarantee that the overall
failure probability will be bounded by $2^{-S}$.}.

\section{Quantum Authentication}\label{sec:quanauth}



In our construction of Multiparty Quantum Computation (MPQC) we
use a quantum authentication scheme (QAS), such as that proposed
in \cite{BCGST02}.
Any QAS based on a quantum CSS code can be used, but some later
protocols become simpler when the QAS is based on a self-dual
code. Therefore our first contribution is a family of self-dual
Quantum Authentication Schemes. We will build a scheme which is
exponentially secure in some arbitrary security parameter $m = 2d
+ 1$, where $d$ is a parameter of the code. The scheme is not
optimal in terms of redundancy or error probability, but it is
sufficient for our purposes.

When using this scheme in the computation, we will assume that $m$
is larger than the security parameter of the Multiparty
Computation times the number of players.
Let $p$ be a prime, $m < p < 2m$. All the algorithms we propose
will manipulate qudits in $\mathbb{Z}_{p}$.

The scheme will be based on a classical key which will be composed
of two parts: $k_1, \ldots, k_m \in_{R} \{ \pm 1\}$ and a string
$x \in_R \{0,1\}^{2m\log_2(p) }$. The dealer will then apply two
transformations:

First, in a way quite similar to the stretched polynomial code,
the dealer will apply
$$\ket{S_a} \rightarrow p^{-d/2}\mathop{\sum_{{deg(f) \le d \atop f(0)=a}}} \ket{k_1 \times f(\alpha_1), \ldots, k_m
\times f(\alpha_m)},$$
where $\alpha_1 , \ldots , \alpha_m \in\mathbb{Z}_{p}$ are
distinct nonzero points known to all players.

Then, very much like the Quantum Authentication Code of
\cite{BCGST02}, the dealer will encrypt the state by applying a
random Pauli operation on each part of the state. This will create
a stretched and shifted polynomial-like code. The encryption will
be denoted as $E_x$, so we will have $\psi = E_x A_k (\phi)$.

Note that the authentication (without the encryption) is
self-dual. To see this, we apply the Fourier transform on the code
transversely as in \cite{AB99}, getting
$Fourier(\ket{S_a})=p^{-1/2}\sum_{b} \omega^{ab} \ket{S_b}.$
This gives:
$$\ket{S'_b} = p^{-d/2}\sum_{{deg(f) \le d \atop  f(0)=b}} \ket{k_1^{-1} \times f(\alpha_1) , \ldots,  k_m^{-1}
\times f(\alpha_m)},$$
which is equal to $\ket{S_{b}}$, as $k_i^{-1} = k_i$. The combined
code $E_x A_k$ remains self dual --- the Fourier transform on
$E_x$ is equivalent to a change in the classical key $x$.

\paragraph{Security of the Quantum Authentication Scheme}

After the encoding, an adversary can try to tamper with the
information as she likes, but without knowledge of $(x,k)=K$.

Finally the receiver takes as input the system $\rho'$, and tries
to return to the encoded state based on $k,x$. We will apply
definitions 1 and 2 of \cite{BCGST02}, and say that the receiver's
output lies in a Hilbert space $M \otimes V$, where M has a size
of $m$ qubits (the size of the original state) and $V$ is a
Hilbert space of dimension 2, with basis states
$\ket{\acc},\ket{\rej }$. Define projectors
\begin{eqnarray*}
P_1^{\ket{\psi}} &=& \ket{\psi} \bra{\psi} \otimes I_V + I_M
\otimes \ket{\rej } \bra{\rej } \\
&&- \ket{\psi} \bra{\psi} \otimes \ket{\rej }\bra{\rej } \\
 P_0^{\ket{\psi}} &=& (I_M -
\ket{\psi} \bra{\psi}) \otimes \ket{\acc} \bra{\acc}
\end{eqnarray*}

\begin{lem}\label{auth-good-lem}


In expectation over $k$ and $x$,
for any encoded state and for any action taken by the adversary,
the receiver's output has exponentially good fidelity to the space
spanned by $P_1^{\ket{\psi}}$ \footnote{In particular, with
exponentially good probability over $k,x$, the fidelity is
exponentially close to 1.}.
If the adversary did not change the authenticated state the output
will be the original state tensored with $\ket{\acc}$.
\end{lem}

\noindent The proof of the lemma appears in \ifnum\ieeefinal=1 the
final version. \else appendix~\ref{app:auth-lem}. \fi

Actually, this security definition is not quite sufficient for our
purposes, since we need the authentication to remain secure in a
variety of contexts.  We adopt the Universally Composable
definition of \cite{HLM} with a $TTP$: That is, the sender passes
the state to the $TTP$.  The adversary then gets to decide whether
the $TTP$ gives the correct state to the receiver, or instead the
state $\ket{\rej }$.  Hayden \textsl{et al.}~\cite{HLM} show that
the class of QAS described in \cite{BCGST02}, including the code
above, remain secure with respect to this stronger definition.

We also need the authentication scheme to remain secure when it is
applied to many states authenticated with the same $k$ but
different $x$'s. The proof is essentially the same: we can treat
the combined system as a single large authentication scheme which
fails if even one of the states fails the authentication test. It
is again sufficient to consider attacks where the adversary
applies a Pauli matrix, and the argument above shows that she is
likely to be caught if she attacks even a single state in this
way.

\paragraph{Operations on Authenticated Quantum Data}

A key advantage of the code we present is that it is possible to
perform Clifford operations on coded data when one party holds the
classical keys and the other party holds the data.
\ifnum\ieeefinal=0 In appendix~\ref{app:clifford}, we show how to
perform these operations (roughly, one applies the Clifford gate
transversally to the encoded state, and then a corresponding
linear operator to the keys). \else In all cases, we can do this
by performing transversal operations on the quantum state (that
is, separate quantum gates on each share), and some corresponding
transformation of the encryption key $x$. For instance, the
Fourier transform can be done applying the Fourier transform
transversally and changing the key $x = (x_0, x_1)$ to $x' = (x_1,
x_0)$, and the $\bmod p$ SUM gate by transversal SUM gates while
transforming the keys $x=(x_0, x_1)$, $y = (y_0, y_1)$ to $x' =
(x_0, x_1 - y_1)$, $y' = (x_0 + y_0, y_1)$. Measurement can be
performed by measuring each qudit to get a classical word, which
can be decoded with the help of the key.\fi

A few properties of the code are:

\begin{enumerate}
\item The new states are still correctly authenticated.

\item Only the correct gate leaves the states correctly
authenticated (since a different gate would require a different
transformation of $x$).

\item Performing any of these operations does not give any new data on the
keys. This is important in the case of CNOT, where knowing the key
$x'$ of one of the states after the operation does not give any
information on the key $y'$ of the other state.

%
%

\end{enumerate}

Measuring an authenticated state according to the standard basis
yields a random codeword which is multiplied and shifted. Changes
to the codeword without knowledge of the keys is equivalent to
applying $X$ operations on the quantum state, and has the same
probability of getting caught.

\paragraph{Handling keys}

In the following sections we will want to manipulate the classical
keys in many ways. We will use an imaginary classical Trusted
Third Party, which implements classical multiparty computation.
From now on, all classical keys of authenticated data will be sent
immediately to this Third Party, which will tell the players the
meaning of their actions based on those keys. While such a $TTP$
does not exist, we can simulate it using (for example) the
classical multiparty computation of \cite{RB89}.

\section{Verified Quantum State Authentication}\label{sec:shareEpr}

In this section, we force the dealer to send each honest player a
correctly authenticated message, using the QAS of section
\ref{sec:quanauth}. A dealer who does not comply will be revealed
as faulty in front of all the players and is kicked out. In the
first subsection, we show how to force the dealer to send
correctly authenticated zero states to every honest player. We
later transform the zero states to EPR pairs shared between
players, and then pass other states using quantum teleportation.
The algorithms succeed only with high probability.



\paragraph{Sending ``Verifiable'' Authenticated Zeros}

Let $D$ be a dealer, who should send states of the form
$Auth_{(k,x)}(\ket{0})$ to all players. The problem is that later
in the protocol, players are required to present states which have
been authenticated by the dealer.  There would then be no way to
distinguish between an honest player who was originally given bad
states by a faulty dealer and a corrupted player who changed the
states she received from an honest dealer.
To solve this problem we incorporate the following protocol, which
guarantees that with exponentially good probability either the
dealer is caught, or every honest player has a large number of
$\ket{0}$ states authenticated by the dealer:

\begin{myprot}

Protocol Zero-Share (Dealer $D$, $r$-copies to each player)

\begin{enumerate}

\item $D$ chooses one random key $k$, and creates many states of
the form $Auth_{(k,x)}(\ket{0})$ for many different $x$'s (all
$x$'s are chosen at random).

\item $D$ sends each player many ($(r + 2s) (t + 1)$) such states.

\item $D$ sends all the keys to the classical $TTP$.

\item Each player $P_i$ performs purification on his states. A
purity testing protocol for zeros which spends $2s$ states is
given later. The results of the measurements are sent to the
$TTP$.

\item The $TTP$ returns each player a bit indicating whether
everything was alright with her states. This gives
fidelity\footnote{In calculating the fidelity we assume that all
the authentication checks succeeded.} of $1 - p^{-s}$ to the
statement that either $P_i$'s states are authenticated zeros or
the dealer is caught by $P_i$\footnote{A more formal definition
can be cast by letting $P_i$ output $\ket{\acc},\ket{\rej }$ as
before and then we have high fidelity to the state of
authenticated zeros tensored with $\ket{\acc}$ or anything else
tensored with $\ket{\rej }$}.

\item For each player $i$, if $P_i$ caught the dealer she
complains about him. If there are more than $t$ complaints the
dealer is faulty.

\item Each player who did not complain distributes $r + 2s$ of
her states to each player who did complain.

\item Each complainer $P_i$ does the following: for all $j$, using
the zero purity test protocol and with the help of the $TTP$ (as
before $P_i$ only measures and the $TTP$ tells him if the states
are good), go over the states you got from $P_j$. Find a player
$P_j$ who you can trust (i.e., $P_j$ gave $P_i$ states which were
correctly authenticated by the dealer).
\end{enumerate}

\end{myprot}

\medskip

\begin{lem}
In the last stage of the protocol, with high probability every
honest player will have zero states which were authenticated by
the dealer.
\end{lem}

\begin{proof}
At most $t$ players complained about the dealer in step 6. This
means that at least one honest player got authenticated zeros, and
she will pass them to all the complainers. Therefore, if after the
last step a player complains that she doesn't have any
authenticated states she is faulty.
\end{proof}

\begin{lem}
The adversary has no new information about the key $k$ used by the
dealer or about the $x$'s in the surviving zero states.
\end{lem}

\begin{proof}
For all $i$, all the measurements made by $P_i$ give random
values, and players are only told by the $TTP$ that the check
succeeded.
\end{proof}

Note that the protocol has an exponentially low probability to
fail completely. This could happen (for example) if the dealer
sends non-zero states, but a single honest player is fooled by the
dealer. This will mean that the dealer will not be considered
faulty in step 6, and all the honest players will fail in the last
step.

The protocol requires a method of testing that a set of states are
(close to) correctly authenticated zeros.   We present such a zero
purity test in appendix~\ref{app:zero-test}.



\paragraph{Generating Authenticated EPR pairs}

To share an authenticated EPR pair with the dealer, $P_i$ takes
two authenticated zeros, and using the classic $TTP$, performs a
transversal Fourier on one of them, and then a SUM. $P_i$ then
sends $D$ one half of the pair.

In order to see the security of the protocol we need to look at
two cases:

\begin{enumerate}

\item $P_i$ is honest but the dealer is faulty: $P_i$ holds zeros
which were authenticated by the dealer (as the dealer was not
kicked out of the protocol Zero-Share). The rest of the protocol
depends on $P_i$.

\item The dealer is honest but $P_i$ is faulty: The EPR pairs
which are authenticated by the dealer will be used to pass
information from the dealer to $P_i$ using quantum teleportation.
A faulty $P_i$ could send the dealer a state which is not part of
an EPR pair (say by destroying the other half or passing some
junk), but this does not add cheating power, as it is equivalent
to destroying the data the dealer is trying to give to
$P_i$\footnote{As we saw, the dealer can not fail this protocol.
Therefore, after using it, $P_i$ will be considered responsible if
anything goes wrong. This alone makes the protocol secure against
a faulty $P_i$ and an honest dealer.}.

\end{enumerate}

\section{Weak Quantum Secret Sharing (WQSS)}\label{sec:WQSS}

Weak Secret Sharing \cite{Rab94, RB89} is a protocol with two
phases.
It provides similar functionality to a classical string commitment
scheme, replacing computational assumptions with the cooperation
of the honest majority.
In the first phase, the dealer shares a (quantum) state (the
secret), among all the players, such that the faulty players have
no information about the state. In the second phase, the (quantum)
data is sent to a reconstructor (sometimes called the receiver),
who reconstructs the secret. We demand that if the dealer is
honest the reconstructor can reconstruct the secret shared. If the
dealer is faulty during the sharing phase some state must still be
determined. However, if the dealer is faulty during the
reconstruction phase the bad players can make sure that no state
is reconstructed. In the case where no state is reconstructed the
reconstructor will know that the dealer is faulty. At the end of
the sharing phase, the players have a state encoded in the quantum
error-tolerant secret sharing scheme of \cite{CGS05}, but with an
additional security guarantee in the case where the dealer is
faulty. As before, we only want our protocol to succeed with high
probability.


We give a formal definition using a $TTP$:

\begin{enumerate}

\item The dealer $D$ sends $TTP$ a state $\rho$, or no state at
all. If D did not send a state the $TTP$ notifies all the players
that this is the case and the protocol ends.

\item Otherwise, at the reconstruction phase, a reconstructor $R$
is chosen.

\item If $D$ is honest, the $TTP$ sends the state $\rho$ to $R$.
If $D$ is faulty, she can tell the $TTP$ not to send the state. In
this case the $TTP$ tells the reconstructor that $D$ is faulty.

\end{enumerate}

The difference between this variant and Verifiable Secret Sharing
lies in step 3 of the definition, where $D$ has a chance to ruin a
previously shared state.

\paragraph{Protocol}


Before the protocol starts (this will be a prerequisite to all our
protocols from now on) we assume that the dealer has some secret
authentication key $k_{dealer}$. The dealer will use this
authentication key with many random encryption keys.

We maintain the invariant that an honest player will never (with
high probability) think that another honest player is faulty.
Therefore if more than $t$ players blame the dealer, she is truly
faulty and can be kicked out of the protocol.

We give the detailed sharing protocol for WQSS in
appendix~\ref{app:WQSS}.  The outline of the procedure is as
follows:

\begin{enumerate}

\item The dealer encodes a number of zero states using a quantum
polynomial code, and transmits the shares to the players using the
authenticated channel from section \ref{sec:shareEpr}.

\item The players and the classical TTP collectively test that the
states are zeros, and are correctly encoded via transversal random
sums.

\item The players use the shared zero states to create a shared
EPR pair.  Half of it is returned to the dealer, who decodes it
and teleports his state through the EPR pair.

\end{enumerate}




The protocols we present later use the WQSS but will never use its
reconstruction. Therefore we present a naive protocol which relies
on revealing the keys. Reconstructing the secret can be made by
sending all the quantum data to the reconstructor, as well as the
key $k$ and the relevant computed $x$ keys. The reconstructor will
open up the authentication of all the states, measure the second
qubit ($\ket{\acc}$ or $\ket{\rej }$) and use only the correctly
authenticated points. If they are {\bf all} from a degree $t$
polynomial code she will reconstruct $\ket{\psi}$; otherwise the
dealer is faulty. This sort of reconstruction, however, spoils the
secrecy of key $k_{dealer}$, which we need throughout the entire
quantum computation. Therefore we will not use the WQSS as
presented here. Still, for completeness of the paper we
temporarily assume this naive reconstruction, and discuss the
security of the protocol. We also assume the reconstructor is
honest, because she gets all the data anyway.

\begin{lem} \label{wqss:honest-dealer}
If the dealer is honest, the protocol is secure. \end{lem}

\begin{proof}
We first prove that the protocol works, and then prove the secrecy
of the data. As the dealer is honest, encoded zeros are being sent
by the established authenticated channels. With high probability,
all the measurement results which are sent to the $TTP$ are either
the right results or they will be discarded (because of the
authenticated channels). This means that with high probability an
honest dealer will pass the test done by the classical $TTP$.

In the final step, the dealer will get a state encoding half an
EPR pair, and she can decode as there are at least $t+1$ correctly
authenticated shares (given by the honest players). The honest
dealer can then transmit her qubit.

Reconstruction of the secret is possible, when considering the
initial encoding with a $t$ degree polynomial as an erasure code,
and discarding shares which are not correctly
authenticated\footnote{Formally, the first thing the receiver is
doing is to open the authentication using the help of the classic
$TTP$. The receiver then measures the last qubit ($\ket{\acc}$ or
$\ket{\rej }$).} as these shares came from bad players. The
reconstructor has at least $t+1$ points which define the secret
(the points held by the honest players), so she can retrieve the
original state\footnote{Actually the reconstructed state has
exponentially good fidelity to the original state tensored with
$\ket{\acc}$.}. Secrecy follows from the no-cloning theorem and
the ability of the reconstructor to reconstruct the right state.
See \cite{CGS05} for a more complete proof of secure
reconstruction.
%
%
%
%
%
%
%
%
\end{proof}

%
%
%

\begin{lem} \label{wqss:faulty-dealer}
If the dealer is faulty, the protocol is secure. \end{lem}

\begin{proof}

Security in this context only means that after the sharing phase
the state has been set, and can no longer be changed by the
dealer. This means (for example) that the adversary knows the
secret (he can choose it in the beginning of the protocol) and the
only thing we should actually take care of is that the adversary
will not be able to change the secret after the sharing phase
(although he is allowed to prevent its decoding). To see that this
is the case we follow the paths of the shares held by the players
who are honest in the reconstruction phase. If the dealer passes
the purity test done by the classical $TTP$, the two shared states
have high fidelity to shared zeros. Therefore, in step 5 (the last
step of the preparation), the honest players hold a state which
has a high fidelity to an EPR-half encoded in a degree $t$ quantum
polynomial code. Whether or not the dealer teleports a quantum
state, there is an invariant: namely, that the states the honest
players hold form $t+1$ points of some degree $t$ polynomial code,
where each such point is authenticated by the dealer. (Note that
if, for example, the dealer measures her half of the EPR pair, the
state collapses, but we still have a polynomial code encoding a
classical state where all the points are authenticated.)
Therefore, in the reconstruction phase there are only two things
which can happen:

1.  The state reconstructed is the state encoded by the honest
players' shares of the polynomial code, which was established in
the sharing phase.

2.  An authenticated share which does not sit on this polynomial
code appears. In this case, w.h.p.\ no secret will be
reconstructed (as there is no degree $t$ polynomial code which
fits all the authenticated points) and the reconstructor knows
that the dealer is faulty.
%
\end{proof}

\section{Verifiable Quantum Secret Sharing}\label{sec:VQSS}

Verifiable Quantum Secret Sharing is also a protocol with two
phases. In the first phase, the dealer shares a quantum state (the
secret) among all the players, such that the faulty players have
no information about the state. In the second phase, the quantum
data is sent to a reconstructor, who reconstructs the secret. We
demand that the value which the reconstructor reconstructs is set
during the sharing phase of the algorithm. As before, we only want
our protocol to succeed with high probability.

The main difference between VQSS and WQSS is the dealer's
capability to ruin the secret after it has been shared. Our main
technique in solving this problem is based on the 2-Good-Trees of
\cite{CCD88,CGS02} or the VQSS of \cite{RB89}. The idea is to
share a secret using a WQSS, and then share each one of the shares
using WQSS. This means that the faulty players no longer have
control of their shares.  They can eliminate their shares by
causing the WQSS reconstruction to fail, but cannot change them to
some other state which could spoil the dealer's original state.
Note that one of the last steps of the protocol we present, in
which the authentication of the first layer is removed, is only
needed for the full MPQC protocol.


We give a definition of Verifiable Quantum Secret Sharing using a
$TTP$:

\begin{enumerate}

\item The dealer $D$ sends $TTP$ a state $\rho$, or no state at
all. If D did not send a state the $TTP$ notifies all the players
that this is the case and the protocol ends.

\item Otherwise, at the reconstruction phase, a reconstructor $R$
is chosen and the $TTP$ sends her the state.

\end{enumerate}

\paragraph{Protocol}

As in WQSS, the sharing phase will have a long preparation part
and then a simple sharing part. During the preparation the dealer
will use a temporary authentication key $k_{dealer}$ in addition
to the standard authentication channels we've established. As a
preliminary step to the algorithm an authenticated channel is
created with this key, and the key will be revealed to all the
players at the end of the preparation. To sum up, the protocol for
VQSS will demand two kinds of secret authentication keys:

\begin{enumerate}

\item The dealer has a temporary key $k_{dealer}$ for the first
level of the tree. When sharing more than one secret in the MPQC
each secret will have a new random key.

\item Each player $P_i$ will have a constant secret authentication
key $k^i$ for the second level of the tree. These keys will be
constant throughout MPQC.

\end{enumerate}

In addition, each authentication has a random encryption key $x$
associated with it, as usual a different one for each state.



We give the detailed VQSS protocol in appendix~\ref{app:VQSS}.
Briefly, it consists of the following steps:

\begin{enumerate}

\item The dealer shares encoded zero states using WQSS, and then
each player further shares the state he receives, again using
WQSS. The players and the classical TTP collectively check whether
the states have been correctly shared and if they are in fact
zeros.

\item The players use the shared zero states to create a shared
EPR pair.  Half of it is decoded and returned to the dealer.

\item Remove the top-level authentication (which uses the
temporary key $k_{dealer}$) from the EPR pair using transversal
Clifford group operations.

\item The dealer teleports his state through the EPR pair.

\end{enumerate}



We note that it is possible to perform transversal Clifford
operations between two shared secrets. The proof of this is very
similar to the possibility of performing transversal Clifford
operations on coded states. The only subtle point is that secrets
shared by different dealers are actually protected in the same way
(as we remove $k_{dealer}$).


We describe the reconstruction protocol here, and defer the
security proof of both protocols to appendix~\ref{app:VQSS}. Like
the sharing protocol the reconstruction uses a one-time
authentication key $k_{reconstructor}$ (with appropriate
authenticated quantum channels), and the same player keys $k^i$.

\begin{myprot}

Protocol Reconstruct-VQSS (Reconstructor R, Key for reconstruction
$k_{reconstructor}$, Player Keys $k^i$,  $1 \le i \le n$)

\begin{enumerate}

\item Reconstructor: Create an EPR pair and share it as in
Share-VQSS. This includes a tree of height 2 and taking out the
top level authentication, but excludes the final teleportation
step.

\item All players and Reconstructor: Use quantum teleportation on
the previously shared secret and on the reconstructor's shared EPR
pair half to transfer the secret to the reconstructor's EPR-half
which is still held by her. This is possible as after the removal
of $k_{reconstructor}$ the codes of the dealer and reconstructor
are actually identical, and it thus is possible to perform
Clifford operations between their secrets.

\end{enumerate}
\end{myprot}

\medskip


Again we begin by assuming that the reconstructor is honest. We
have two lemmas which together prove the security of the combined
sharing-reconstruction protocol.

\begin{lem} \label{honest-dealer-vqss}
If the dealer is honest, with exponentially good probability the
faulty players cannot affect the reconstruction of the secret.
Moreover, no player but the receiver learns anything (in the
information theoretic sense) about the secret.
\end{lem}

\begin{lem} \label{faulty-dealer-vqss}
If the dealer is faulty, with exponentially good probability he
can not change the secret he shared. Moreover the faulty players
do not learn $k^i$ values for honest players.
\end{lem}

\noindent Both lemmas are proved in appendix~\ref{app:VQSS}.

If the reconstructor is faulty, the only secret we are trying to
protect is the player keys $k^i$. Their security stems from the
security of performing Clifford group operations on coded states.

\section{Multiparty Quantum Computation}\label{sec:MPQC}

Our VQSS scheme already resembles Multiparty Quantum Computation
in the ability to share a few secrets in parallel (all with the
same player authentication keys $k^i$), and use Clifford
operations between them. In order to complete this to a Multiparty
Quantum Computation we need to add a Toffoli gate. We base the
creation of the Toffoli gate on a shared Toffoli state a la
\cite{Sho96}, and the sharing of our Toffoli state on the ideas of
the Power-Tables in \cite{RB89}.

Let $\frac{1}{p} \sum_{a,b} \ket{a,b,ab}$ be the Toffoli State.
\ifnum\ieeefinal=1 It is well-known \else In
appendix~\ref{app:Toffoli-gate} we show \fi
that it is possible to perform a Toffoli on any state using
Clifford operations and this state.

Sharing the Toffoli state can be done by our protocol for VQSS. We
begin by sharing many Toffoli states which are only polynomially
good\footnote{It might be possible to use Bravyi and Kitaev's
technique in \cite{BK04} and thus obtain exponential fidelity
after passing a constant barrier. However, using their technique
would require us to describe the entire calculation in $F_{2^q}$
instead of in $Z_p$ as we did here.} and then use a fault-tolerant
multiparty computation (for example \cite{AB99}) to create an
exponentially good Toffoli state from an encoded zero. The
polynomially good states are created by some arbitrary player. If
the player fails to create polynomially good Toffoli states she is
faulty and is kicked out of the protocol.


\begin{myprot}

Protocol Create-Toffolis (Arbitrary Dealer $P_i$, Player Keys
$k^i$ for $1 \le i \le n$)

\begin{enumerate}

\item $P_i$: Share a polynomially large number of Toffoli states.

\item All players: Run state tomography on all but a polynomial
fraction of the shared states to check the states sent by $P_i$.
If $P_i$ is caught not sending Toffoli states, she is kicked out.
Otherwise, the states sent have a polynomial fidelity to the
Toffoli state.

\item All players: Using the states which were left (many states
were opened up in the previous step), create an error correcting
computation which creates a Toffoli state with exponential
fidelity from the polynomially good Toffoli states. This can be
done using the protocol described below, which uses standard
techniques of noisy computation such as \cite{AB99}.

\end{enumerate}
\end{myprot}

\medskip

\noindent In appendix~\ref{app:purify-Toffoli} we present a
protocol especially designed to purify Toffoli states.

\section{Simulations of the protocol}\label{sec:simul}

Proving the Universal Composability of our algorithm may seem at
first like a daunting task. Surprisingly, this is not the case,
and the simulation turns out to be almost trivial. We sketch the
main details of the simulation in \ifnum\ieeefinal=0
appendix~\ref{app:simulation}. \else the full version. \fi Here we
simply note a few key features which we use:

\begin{enumerate}

\item We rely heavily on the universal composability of classical
multiparty computation, and therefore use an ideal Classical
Trusted Third Party (called here $C-TTP$ to stress that it's
classical) in our simulation.

\item The preparation step of the protocol is generic and
independent of the inputs (and even the function we evaluate).
This makes its simulation trivial.

\item Using teleportation enables us to avoid passing quantum data
in the input and output phases of the protocol. Instead, the
players only pass (and obtain) classical data to (and from) the
$C-TTP$. Moreover, this data is uniformly distributed and
independent of the quantum data in the computation.

\item For any group containing $\le t$ players, the results of any
transversal measurement done during the computation are random,
uniformly distributed, and independent of the encoded quantum
state which is being measured.

\end{enumerate}



%

\section{Acknowledgments}\label{sec:ack}
M.B.~and A.H.~were partially supported by an Israel Science
Foundation research grant and by an Israel Ministry of Defense
research grant.  D.G.~is supported by CIAR and NSERC of Canada.


\appendix


\ifnum\ieeefinal=0 
\section{Proof of Lemma \ref{auth-good-lem}}
\label{app:auth-lem}

\noindent {\bf Lemma~\ref{auth-good-lem}.} {\it In expectation
over $k$ and $x$, for any encoded state and for any action taken
by the adversary, the receiver's output has exponentially good
fidelity to the space spanned by $P_1^{\ket{\psi}}$. If the
adversary did not change the authenticated state the output will
be the original state tensored with $\ket{\acc}$.}

\begin{proof}

Assume the adversary operated on the coded word. By
\cite{BCGST02,HLM}, it is enough to consider Pauli operations done
by the adversary, and if we can catch each such operation with
exponentially good probability, then the code is secure.  We can
view such an attack as adding some set of values (mod $p$) to the
qudits of the code in the computational basis and another set in
the Fourier-transformed basis.

Note that if the adversary applies a non-identity Pauli operation
which acts on at most $d$ qudits he will be caught (that is, the
output on $V$ is going to be $\ket{\rej }$) with probability $1$,
as the code can detect up to $d$ errors. If the adversary applies
a Pauli operation which acts non-trivially on more than $d$
qubits, the adversary succeeds only if he adds a signed degree
$\leq d$ polynomial in each basis, with the correct sign factors
$k_i$.

To see that he cannot reliably do this, assume without loss of
generality that $\alpha_{1}, \ldots ,\alpha_{r}$ are the points
which were not changed by the adversary, $r \le d+1$, and suppose
that he adds $\ket{\beta_1, \ldots, \beta_m}$ in one of the bases
(so $\beta_i = 0$ for $i \leq r$ and $\beta_i \neq 0$ for $i >
r$). There will be exactly one polynomial $f$ of degree at most
$d$ which is consistent with the equations $k_i f(\alpha_r) =
\beta_i$ for $i \leq d+1$.  The probability that any further
non-zero point $\beta_i$ ($i > d+1$) is consistent with this
polynomial is at most $1/2$, as it will be inconsistent with at
least one of $k_i = +1$ or $k_i = -1$. Therefore the probability
that the adversary is detected is at least $1 - 2^{-d}$.

The second part of the claim is trivial --- if the adversary
applies an identity the original state will be reconstructed.

\end{proof}
\fi 


\ifnum\ieeefinal=0 
\section{Clifford Operations on Encoded States}
\label{app:clifford}

In this appendix we describe how to perform Clifford operations on
encoded states. Similar techniques for gates on encrypted states
were used in \cite{Chi05}.

The basic setting here is again a dealer who authenticates the
data and holds the keys, an adversary who holds the quantum data
and a receiver which at the final stage of the protocol gets the
data as well as the keys and outputs a state in $M \otimes V$.


Applying a Pauli operation on the encoded state is equivalent to
applying a transversal Pauli operation on the encoding of the
state, which is equivalent to a shift in the classical vector $x$
where the key is $k,x$. In a very similar manner, multiplying the
encoded state with a constant also maps to a change in $x$.

Measurement in the computational basis is straightforward.
Measuring each qudit results in a classical word, encrypted using
part of the key $x$, of the form $(k_1 \times f(\alpha_1), \ldots,
k_m \times f(\alpha_m))$, with $f$ a polynomial of degree at most
$d$, random except for $f(0)$, which tells us the decoded outcome
of the measurement.

Applying the Fourier transform on encoded states is also possible
because the code is self-dual.  We apply the transversal Fourier
transform as above, and adjust the encryption key $x$
appropriately.  In particular, $x = (x_0, x_1)$, giving shifts in
the computational and Fourier bases, must be adjusted to $x' =
(x_1, x_0)$.  We can also perform the phase gate $\ket{j}
\rightarrow \omega^{j(j-1)/2} \ket{j}$ transversally, although we
will not actually need it in any of our circuits.

Finally, the code enables us to perform the $\bmod p$ SUM
operation between states authenticated by the same $k$, with
different $x$ keys. Let $\psi$ be a state authenticated using $k$
and $x$, and $\phi$ using $k$ and $y$. Then marking the
authentication with $A_k$ and the encryption by $E_x$ and $E_y$, a
transversal SUM on $A_k(\psi)$ and $A_k(\phi)$ maps to SUM on the
data, but with new encryption keys $E_{x'}$ and $E_{y'}$. The
transformed encryption keys can be determined using the Gottesman
Knill algorithm (see \cite{NC00}).
\begin{eqnarray*}
E_x(A_k(\ket{\psi})) + E_y (A_k(\ket{\phi})) =
\\ (E_{x'} \otimes E_{y'})
[A_k(\ket{\psi}) + A_k(\ket{\phi})] =
\\ (E_{x'}A_k \otimes E_{y'}A_k)
(\ket{\psi + \phi})
\end{eqnarray*}
In particular, $x=(x_0,x_1)$, $y=(y_0,y_1)$ become $x' = (x_0, x_1
- y_1)$ and $y' = (x_0 + y_0, y_1)$.

\fi 


\section{A Zero Purity Test}
\label{app:zero-test}

We describe a simple zero purity test for states $\ket{\phi_0},
\ldots \ket{\phi_w}$ (for some $w$), to test if they are all
authenticated correctly and are encoding the qudit $\ket{0}$.
Choose $a_0, \ldots a_w \in_R \{0,1, \ldots ,p\}$ and calculate
the transformation $\sum a_i \ket{\phi_i} \rightarrow
\ket{\phi_0}$ where the sum is done by using SUM. Then, open up
(i.e., measure) the new state $\ket{\phi_0}$. Correct the state
you have using the keys, applying shifts and multiplications. You
should be holding a polynomial of degree $\le d$ with free
coefficient $0$. Run the same check (with new random numbers and
without $\ket{\phi_0}$ which was already spent) on the Fourier
transform of the states, although now the free coefficient is
random. Iterating this $s$ times spends $2s$ states and gives the
desired fidelity. All the operations (multiplying with constants,
SUM's and measurements) are done with the help of the classical
$TTP$, as only the $TTP$ holds the classical keys.

In a way similar to the definitions of authenticated states, let
$M$ be the Hilbert space which holds the remaining states
$\ket{\phi_0} , \ldots , \ket{\phi_{w-2s}}$, let $\ket{\psi} =
\ket{Auth(0)} \otimes \ket{Auth(0)} \otimes \cdots \otimes
\ket{Auth(0)} \in M$, and let V be a Hilbert space of dimension 2,
with basis states $\ket{\acc},\ket{\rej }$. Define projectors
$$P_1^{\ket{\psi}} = \ket{\psi} \bra{\psi}
\otimes I_V + I_M \otimes \ket{\rej } \bra{\rej } - \ket{\psi}
\bra{\psi} \otimes  \ket{\rej } \bra{\rej }$$
$$P_0^{\ket{\psi}} = (I_M - \ket{\psi} \bra{\psi})
\otimes \ket{\acc} \bra{\acc}$$

\begin{lem}

The result of the zero purity test has fidelity $1 - O(p^{-s})$ to
the space spanned by $P_1^{\ket{\psi}}$. If the adversary did not
change the authenticated state the output will be $\ket{\psi}
\otimes \ket{\acc}$.

\end{lem}

\ifnum\ieeefinal=1 The proof of the lemma is deferred to the final
version. \else

\begin{proof}

Again according to \cite{CGS02} and \cite{BCGST02} it is enough to
see that with very high probability our check finds Pauli
operations on the encoded states. (Note that we assume the
authentication goes well and only discuss the encoded data.) The
first check finds an $X$ operation with probability $1 - 1/p$, and
the second one finds a $Z$ operation with the same probability. As
the checks commute, iterating them $s$ times finds any Pauli
operation with probability $1 - p^{-s}$ assuming all
authentication tests succeeded.
\end{proof}
\fi


\section{Weak Quantum Secret Sharing}
\label{app:WQSS}

In this appendix, we give the detailed protocol for sharing states
using WQSS.

\medskip

\begin{myprot}

Protocol Weak-Quantum-Secret-Sharing (Dealer D, Dealer Key
$k_{dealer}$)\footnote{Underlined and words are the name of the
phase. The phase continues until a new phase begins.}

\begin{enumerate}

\item {\it \underline{Preparation}} Dealer $D$: Encode many ($2s +
2$, where s is the security parameter) zeros using the quantum
polynomial code of degree $t$ and length $n$. For each zero
encoded send the $i$th share to $P_i$, using the authenticated
channel established in section \ref{sec:shareEpr}. Note that the
authenticated channel is used with the same $k_{dealer}$ all the
time, but with different $x$'s.

\item All players and classical-$TTP$: Using random numbers
generated by the classical $TTP$, the players perform transversal
random sums, both in the standard and the Fourier basis. The
players measure $2s$ of their shares ($s$ checks in each basis)
and send the results to the classical $TTP$.

\item Classical-$TTP$: Discard values which do not authenticate
correctly. These values must come from bad players (we are using
verified authenticated channels).

\item Classical-$TTP$: If errors are detected in the outer
polynomial code the dealer is faulty. All players are informed and
the protocol is aborted.

\item All players: The players collectively generate an EPR pair
from the two remaining zeros and send one half of the pair to the
dealer.

\item {\it \underline{Sharing}} Dealer: Decode the EPR-half you
have,\footnote{Note that the dealer can perform this step only
because he knows all the keys he used in step 1.} and using
quantum teleportation send the secret, giving your measurement
results to the $TTP$. This results only in a change of keys ---
the players do not need to act on their states or manipulate any
new information.

%

\end{enumerate}
\end{myprot}

\begin{figure}
\epsfig{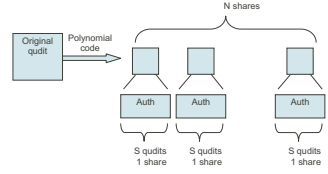}
\caption{Schematic of Weak QSS}
\end{figure}

\section{Details of VQSS}
\label{app:VQSS}

Below is the detailed protocol for sharing in VQSS:

\medskip

\begin{myprot}

Protocol Share-VQSS (Dealer $D$, Secret $\psi$, Key for the secret
$k_{dealer}$, Player Keys $k^i$ for $1 \le i \le n$)

\begin{enumerate}

\item {\it \underline {Preparation Phase}} Dealer: Prepare many
($2s+2$) zeros and encode them with the polynomial code of degree
$t$ and length $n$. Send the $i$'th share ($R_i$) to the $i$'th
player using the authenticated channel using key $k_{dealer}$.

\item Player $P_i$: Take each state shared by the dealer, and
share it using WQSS and your key $k^i$. Mark the $j$'th share as
$R_{i,j}$. Note that for each zero-share $P_i$ got from the
dealer, she has to generate $2s + 2$ new zeros.

\item All players and classical-$TTP$: Perform transversal random
sums in both standard and Fourier basis to check that the zeros
shared by the dealer are OK. All results are sent to the classical
$TTP$. Note that the shares $R_{i,j}$ are being manipulated here,
and that $2s$ of the original zeros are being spent.

\item Classical-$TTP$, operating on the measurement results that
were given by the players: If a measurement result on $R_{i,j}$ is
not authenticated, $P_j$ is faulty and this result should be
ignored. If $R_i$ is not properly reconstructed as a state
authenticated by the dealer, $P_i$ is faulty; inform all players
of this. If the top level does not decode to $\ket{0}$ (when
ignoring faulty $R_i$'s), the dealer is faulty. If the dealer is
faulty the classical $TTP$ tells that to all the players and the
dealer is kicked out.

\item {\it \underline{Generating EPR}} All players $P_j$ with the
help of classical-$TTP$: Using transversal operations on two
shared zeros generate an EPR pair. This is done by acting on
$R_{i,j}$ shares.

\item {\it \underline{Sending the EPR Half to the Dealer}} Player
$P_j$: Send half of the pair created from the $R_{i,j}$'s to
$P_i$.

\item Player $P_i$: Using the shares you received (using only ones
which are correctly authenticated by you) decode $R_i$ and send it
back to the dealer. Note that $P_i$ knows $k^i$, and does not need
the help of the classical $TTP$ here.

\item Dealer: Decode the state you received, discarding any
incorrectly authenticated shares.

\item {\it \underline{Getting Rid of $k_{dealer}$}} Players $P_j$
with the help of classical-$TTP$: Using transversal Clifford
operations remove the top level authentication. This is possible
as each $R_i$ was authenticated and the qudits were shared by
$P_i$. Using transversal operation on these qudits we can decode
$R_i$ and leave this coordinate protected only by WQSS using
$k^i$. This step is not essential for VQSS but only for MPQC.

\item {\it \underline{Sharing}} Dealer: Share your secret using
quantum teleportation, passing measurement results to the
classical $TTP$. This only results in changes of $x$ keys (just
like the sharing in WQSS).

%
%
\end{enumerate}
\end{myprot}

\begin{figure}
\epsfig{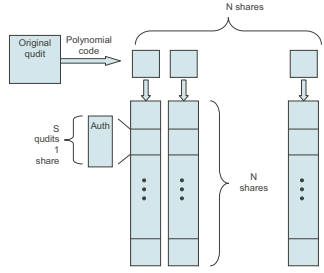}
\caption{Schematic of VQSS}
\end{figure}

\medskip

\ifnum\ieeefinal=0 

 We now sketch proofs of lemmas~\ref{honest-dealer-vqss}
and~\ref{faulty-dealer-vqss}. Again we begin by assuming that the
reconstructor is honest. The two lemmas together prove the
security of the protocol.

{\it  If the dealer is honest, with exponentially good probability
the faulty players cannot affect the reconstruction of the secret.
Moreover, no player but the receiver learns anything (in the
information theoretic sense) about the secret.}

\begin{proof}
The proof of this lemma is very similar to that of lemma
\ref{wqss:honest-dealer}, and is actually quite simple. The only
subtle point is the fact that removing the top level
authentication from the shares given to bad players does not
endanger the secret. This is true however as when the $R_i$ shares
were in the hands of the bad players they were still
authenticated, and when the authentication is removed these shares
are distributed via WQSS and can thus be eliminated by the bad
players but not changed. Eliminating the $R_i$ shares which were
once given to faulty player does not prevent reconstruction, as we
have an erasure code which can correct $t$ erasures.
\end{proof}

{\it If the dealer is faulty, with exponentially good probability
he can not change the secret he shared. Moreover the faulty
players do not learn $k^i$ values for honest players.}

\begin{proof}
The second part of the lemma stems from the security of Clifford
operations on encoded states. As for the first part, if the dealer
passed the tests, the state has a high fidelity to a 2-Good tree
encoding zeros shared by all the players. Creating a transversal
EPR pair from the zeros succeeds because it relies on the shares
on the good players. As in the proof of lemma
\ref{wqss:faulty-dealer}, quantum teleportation would now result
in creating an authenticated state, even if the dealer is faulty.
This state can not be changed --- the coordinated $R_i$ shared by
honest players will be reconstructed correctly, and faulty players
can destroy at most $t$ $R_i$ coordinates, but not change any of
them. As the top level encoding was an erasure code which can
correct $t$ erasures this is ok.
\end{proof}

\fi 

\ifnum\ieeefinal=0 

\section{Performing Toffoli Using Toffoli States}
\label{app:Toffoli-gate}

Assume $\ket{x}$, $\ket{y}$ and $\ket{z}$ were authenticated by
the same player. We will want to perform the Toffoli gate $z
\rightarrow z \oplus x * y$. The problem is that we can only
perform local Clifford operations and SUM's. We will show how to
implement Toffoli with Clifford operations and a known initial
state using a technique from \cite{Sho96,NC00}.

Let $\frac{1}{p} \sum_{a,b} \ket{a,b,ab}$ be the Toffoli State.
Then:

$$ \frac{1}{p} \sum_{a,b} \ket{a,b,ab,x,y,z}  \stackrel {{C_{1}^{-}
-X_4,\ C_{2}^{-} - X_5,\ C_{6}^{+} - X_3}}  {\longrightarrow} $$

$$\frac{1}{p} \sum_{a,b} \ket{a,b,ab + z,x - a,y - b,z}
\stackrel{H_6}{\longrightarrow} $$

$$\frac{1}{p^{\frac{3}{2}}} \sum_{a,b,\ell} \omega^{\ell z}\ket{a,b,ab +
z,x - a,y - b,\ell} \stackrel{Measure (6)}{\longrightarrow} $$

$$\frac{1}{p} \sum_{a,b} \omega^{\ell z}\ket{a,b,ab + z,x - a,y - b,\ell}
\stackrel{C_{1}^{+} - Z_{2}^{\ell},\ Z_{3}^{-\ell},\
Zero(6)}{\longrightarrow} $$

$$ \frac{1}{p}\sum_{a,b} \omega^{\ell z + \ell ab -\ell(ab+z)}\ket{a,b,ab +
z,x - a,y - b,0} \stackrel{Measure(5)}{\longrightarrow}
$$

$$ \frac{1}{\sqrt{p}}\sum_{a}\ket{a,b,ab + z,x - a,y - b,0}
\stackrel{X_{2}^{y-b},\ C_{1}^{+} - X_3^{(y-b)},\ Zero(5),\
Measure(4)}{\longrightarrow} $$

$$ \ket{a,y,ay+z,x-a,0,0} \stackrel{X_{1}^{x -a},\ C_{2}^{+} - X_3^{(x-a)},\ Zero (4)}\longrightarrow
\ket{x,y,xy+z,0,0,0}$$

Performs Toffoli on $\ket{x},\ket{y},\ket{z}$

\fi 


\section{Purifying Toffoli States}
\label{app:purify-Toffoli}

We present an algorithm to purify Toffoli states, which is
interesting in its own right.  Each Toffoli state allows us to
perform one Toffoli gate; a perfect Toffoli state gives a perfect
Toffoli gate, whereas a state with error $\epsilon$ produces a
gate which is $O(\epsilon)$ away from a correct Toffoli gate.
Then we can use techniques of fault-tolerant quantum computation
to turn these noisy gates into an exponentially more reliable one.

Choose some $m$ and $d$, with $m = 3d+1$, $p > m$, $m = O(s)$.

Let $H_{good}$ be the space spanned by $m$ Toffoli states which
were affected by at most $m/8$ non-identity Pauli operations. Then
any $m$ Toffoli states coming from players who have passed the
polynomial state tomography phase will have fidelity $1 -
2^{-O(s)}$ to their projection on $H_{good}$. We now show how to
distill a single exponentially good Toffoli state from the $m$
states.

Let $\beta_{1}, \ldots \beta_{m} \in \mathbb{Z}_p$ be distinct
nonzero points. Look at the state
\begin{multline*}
\sum_{a,b} \sum_{deg(f) \le d \atop f(0) = a} \sum_{deg(g) \le d
\atop g(0) = b} \sum_{deg(h) \le 2d \atop h(0) = 0}
\Big(\ \ket{f(\beta_{1}) , \ldots, f(\beta_{m})} \\
\otimes \ket{ g(\beta_{1}) , \ldots, g(\beta_{m}), h(\beta_{1}) ,
\ldots, h(\beta_{m})}\ \Big).
\end{multline*}
Note that this state can be created exactly by Clifford operations
(which are free in the model we discuss). Now, for $1 \le i \le m$
use a polynomially good Toffoli state (one of the $m$ states we
have left after the first part of the purification) to perform a
Toffoli gate on coordinates $i, m + i, 2m + i$. If the Toffoli
states were ideal this should result in the state $\ket\tau =$
\begin{multline*}
 \sum_{a,b} \sum_{deg(f) \le d \atop f(0) = a}
\sum_{deg(g) \le d \atop g(0) = b} \sum_{deg(h) \le 2d \atop h(0)
= ab} \Big(\ \ket{f(\beta_{1}) , \ldots, f(\beta_{m})}
\\ \otimes  \ket{g(\beta_{1}) ,
\ldots, g(\beta_{m}), h(\beta_{1}) , \ldots, h(\beta_{m})}\ \Big).
\end{multline*}
As the entire $m$ states have exponentially good probability to a
state in $H_{good}$, with exponentially good probability we can
correct $d/2$ mistakes on each one of the three codes, and be in a
state which has exponential good fidelity to $\ket{\tau}$.
Decoding the state using only Clifford group operations we get a
state with exponential fidelity to $\frac{1}{p}
\mathop{\sum_{a,b}} \ket{a,b,ab}$, which is the Toffoli State.

Note that the error rate in this purification step drops from
$\eta$ down to $\eta^{O(d)}$. We can select a (polynomially) large
$d$ or iterate the procedure with smaller $d$ to obtain the
desired fidelity.

\ifnum\ieeefinal=0 
 Note that the procedure presented here could
have worked for any state which is universal (with the Clifford
group) and not just the Toffoli state. This can be done by
replacing the Toffoli's acting on coordinates $i,m+i, 2m+i$ by a
short computation using the universal state which does Toffoli on
three qudits. \fi

\ifnum\ieeefinal=0 

\section{Proof Sketch of Simulations}
\label{app:simulation}

\paragraph{Static Adversary}

With these key points in mind, we can present the simulation of
the protocol, borrowing settings and definitions from \cite{Can01}
and \cite{BM04}. Let $A$ be an adversary in the real protocol.
Given $A$ we define a simulator $S$, which operates together with
the $Q-TTP$ that evaluates the function and simulates $A$'s
behavior. We describe $S$'s behavior in each phase of this
simulation (also called the virtual protocol).

\begin{myprot}

\begin{description}
\item[Preparation Phase:] $S$ runs the protocol simulating all
players, with the faulty players acting according to $A$. The
simulation is exact (and trivial) as no input has been entered by
the honest players. At the end of this phase the simulator holds
shared EPR pairs with every faulty player who was not caught
cheating.

\item[Input Phase:] Bad players who get caught during the
preparation phase in the real protocol do not enter their inputs
to the computation.  Therefore the simulator $S$ will not enter
their inputs to the Quantum Trusted Third Party ($Q-TTP$), which
will inform all players that those players are faulty. For other
bad players, $S$ continues to simulate the teleportation phase,
passing the classical data to the $C-TTP$ computation.

$S$ then decodes each teleported qudit using $t+1$ shares of the
simulated QVSS, and passes it as input to the $Q-TTP$. At this
point $S$ also instructs the honest players to pass their input to
the $Q-TTP$. To continue the simulation $S$ needs to generate the
classical information the honest players would have sent to the
$C-TTP$ during the teleportation, and it does so by generating
uniformly random numbers instead.

\item[Computation:] $S$ performs the simulation by running the bad
players including their interaction with the $C-TTP$ and the
environment. Note that during this phase the players may pass
outcomes of measurements to the $C-TTP$ but never receive any
message back. Inputs to the $C-TTP$ from the honest players are
just random bits.

\item[Output:] The $Q-TTP$ passes the outputs of the honest players
directly to them, and passes the outputs of the bad players to
$S$. To deliver the output to a bad player, $S$ first encodes it
using the QVSS and then teleports it as in the protocol. This
teleportation uses only the coordinates of the $t+1$ virtual good
players to decode the classical data the $C-TTP$ would send the
the faulty player.
\end{description}

\end{myprot}

\paragraph{Adaptive Adversaries}

The simulation for an adaptive adversary is also quite simple. We
only denote the differences between the simulations:

\begin{enumerate}

\item In the input phase: After decoding the input of a bad player
and passing the "real" input to the $Q-TTP$ the simulator (who
holds at least $n-t$ coordinates) can replace the value entered to
be $\ket{0}$ for all players.

\item In the computation phase we simulate all the players (with
input $\ket{0}$). Note that the simulation is exact from the point
of view of the adversary (as it always holds less than $t$ shares)
and therefore the distribution of the players who get corrupted
during the algorithm is correct.

\end{enumerate}

%
%
%
%
%

\fi 

\end{document}